\documentclass[11pt]{article}

\usepackage[margin=1in]{geometry}
\usepackage{titling}
\usepackage{enumitem}
\usepackage{parskip}
\usepackage{booktabs}
\usepackage{array}
\usepackage{graphicx}
\usepackage{float}
\usepackage{caption}
\usepackage{subcaption}
\usepackage[T1]{fontenc}
\usepackage{lmodern}
\usepackage{microtype}
\usepackage{amsmath}
\usepackage{xcolor}
\usepackage{hyperref}

\hypersetup{
    colorlinks=true,
    linkcolor=black,
    urlcolor=blue,
    citecolor=black
}
\usepackage{tikz}
\usetikzlibrary{positioning, arrows.meta, shapes.geometric, fit, calc}

\pretitle{\begin{center}\Large\bfseries}
\posttitle{\end{center}\vspace{-0.5em}}
\preauthor{\begin{center}\normalsize}
\postauthor{\end{center}\vspace{-1em}}
\predate{\begin{center}\small}
\postdate{\end{center}\vspace{-1em}}

\title{Can Humans Detect AI?\\Mining Textual Signals of AI-Assisted Writing\\Under Varying Scrutiny Conditions}
\author{
    Daniel Tabach \\
    Georgia Institute of Technology \\
    \texttt{dtabach3@gatech.edu}
}
\date{April 2026}

\begin{document}
\maketitle
\thispagestyle{empty}

\section*{Abstract}

This study asks whether the \textit{threat} of AI detection changes how people write with AI, and whether other people can tell the difference. I built a two-phase controlled experiment where 21 participants wrote opinion pieces on remote work using an AI chatbot. Half were randomly warned that their submission would be scanned by an AI detection tool. The other half received no warning. Both groups had access to the same chatbot. In Phase~2, 251 independent judges evaluated 1{,}999 paired comparisons, each time choosing which document in the pair was ``written by a human.'' Judges were not told that both writers had access to AI. Across all evaluations, judges selected the warned writer's document as human 54.13\% of the time versus 45.87\% for the unwarned writer. A two sided binomial test rejects chance guessing at $p = 0.000243$, and the result holds across both writing stances. Yet on every measurable text feature I extracted, including AI overlap scores, lexical diversity, sentence structure, and pronoun usage, the two groups were indistinguishable. The judges are picking up on something that feature based methods do not capture.

\section{Introduction}

Large language models have made AI-assisted writing nearly indistinguishable
from human-authored text. This has raised real concerns in academia,
journalism, and the workplace about whether anyone is still writing their
own content or relying on AI tools to write content for them.
Universities for example have responded by adopting AI detection tools. The question covered in this paper will look at whether or not a \textit{threat} of AI-detection actually changes how
participants write with AI, and if it does, can other people tell?

This project centers the question through a two-phase controlled
experiment. I built a Streamlit app that sets up participants with an AI
chatbot and randomly assigns them to one of two conditions. The control
group receives no extra context, while the treatment group is told that their final
submission will be analyzed by an ``AI detection tool.'' Everyone answers the same prompt to form a pair of opinion pieces on remote work, one ``For'' and one
``Against.'' In Phase~2, a separate pool of judges reads the paired documents
(one control, one test, same stance) and has to pick which one they believe was ``written by a human.'' They go through a funneled survey designed in Streamlit that allows them to choose which of our writer's responses were ``written by a human.'' The judges' task was intentionally deceptive. Both groups in the experiment had access to identical chatbots, so judges were not picking ``human vs AI.'' Instead, the response from the judges was testing to see if the judge can identify which writer had been warned that detection software would read the result. The initial hypothesis of this research was that writers who were warned of an AI detection software were more likely to edit and re-frame their writing to sound more ``human,'' and judges would have been able to identify writings in the test group at a significantly higher rate as humans compared to the control group. The hypothesis was intended to measure the incremental behavioral change of an AI-detection threat for the treatment group.

The two research questions are:

\begin{enumerate}[nosep]
    \item \textbf{RQ1 (Behavior under scrutiny):} Do writers behave
          differently when they know their submission will be scanned for
          AI? This covers both interaction metrics (turns, duration,
          tokens, self-reported reliance) and textual features of what
          they submit.
    \item \textbf{RQ2 (Human judges):} Do human judges detect AI
          reliance? Does confidence, reading time, or belief alignment
          moderate their accuracy?
\end{enumerate}

\textbf{Preview of findings:} The initial hypothesis was directionally
supported, but the treatment effect is small. Across n = 1,999 paired evaluations from
251 judges, the warned (test) document was identified as the human written
one 54.13\% of the time, versus 45.87\% for the unwarned (control)
document (95\% Wilson confidence interval on the control as human rate:
[43.70\%, 48.06\%]).
A two sided exact binomial test against the chance
rate of 0.5, treating each response as independent, rejects random
guessing (p = 0.000243 which is significant at alpha < 0.05).
Judges on average rated the warned writers as more human, which is consistent with the
hypothesis, but the difference is only a few percentage points above chance. However, this difference is not insignificant.

The more interesting finding is the disconnect between that signal and
the writing itself. On every behavioral and style feature tested, warned and unwarned writers
were indistinguishable. Both groups leaned on the chatbot
at similar rates: control writers had an AI overlap score of 0.978,
test writers 0.953. The overlap score is an engineered score (0 to 1 scale) that measures how much of a submission matches word for word any AI response the writer saw during the chat in their session (It doesn't consider turns or draft edits - it only looks at how much of the submitted text matched a full AI response in the chat). 1 means the submission is an exact copy-paste from an AI message, and a score near 0 means the writer rewrote the text in their own words or never worked with the chatbot to begin with. The score catches copy-paste moments but its limitation is that it says nothing about what happened in the chat before the paste. For example, two writers can both overlap with the AI exactly, but one might be pasting after asking for a single draft, and another might have iterated with the AI on tone and length for ten minutes and then pasted the final version. The score in this case would look identical. So the judges differences in their selections are not coming from whether or not a writer used or AI or not, but other cues that changed the way they chose the "human" document.

Section~2 covers the experimental design and the decisions made for each step in the experiment. Section~3 covers the platform,
testing methods, and the recruitment methods I used to land participants. Section~4 presents Phase~1 results from the writers (agnostic of the judges evaluations). Section~5 covers the Phase~2
judge evaluation. Section~6 discusses the main takeaways from this study.

\section{Experimental Design}

Every design choice in this experiment was made for a reason. This section
walks through each one and explains the thought process behind each
decision layer.

\subsection{Study Structure}

Each participant sees only one version of the study. They are placed in
either a treatment group (which sees an AI detection label) or a control
group, and they complete two writing tasks: one arguing in favor of remote
work and one arguing against it.

Condition assignment follows an alternating rule based on participant ID.
Odd numbered IDs go to control; even numbered IDs go to test. This is
deterministic rather than random, which is acceptable for a small sample
where true randomization offers limited additional benefit. I wanted to
make sure sample sizes were roughly equivalent across groups. Unit tests
in the codebase confirm that the assignment logic produces exact 50/50
splits on condition for any sample size.

I also rotated the writing order so that half the participants argued in
favor first and half argued against first. This is a standard technique
called counterbalancing, where you spread out order effects so they do
not pile up in one group. If every participant wrote their preferred
stance first, I could not separate ``wrote better because they believe
it'' from ``wrote better because it was the first task and they were
fresher.'' The rotation follows a four-participant cycle that repeats for
the full sample (Brooks, 2012). Table~\ref{tab:assignment} shows how this
works in practice. The drawback is that it does make the sample slightly
noisier to read.

\begin{table}[H]
\centering
\small
\caption{Assignment logic. The four-participant cycle repeats indefinitely for equal session counts across conditions and writing orders.}
\label{tab:assignment}
\begin{tabular}{@{} c l l l @{}}
\toprule
\textbf{PID} & \textbf{Condition} & \textbf{Task 1} & \textbf{Task 2} \\
\midrule
1 & Control & For    & Against \\
2 & Test    & For    & Against \\
3 & Control & Against & For \\
4 & Test    & Against & For \\
\bottomrule
\end{tabular}
\end{table}

One limitation worth noting: for simplicity, I did not implement
persistent sessions. If a participant dropped out due to internet or
browser issues, their session would unfortunately be incomplete, and
indeed this did happen. This was a deliberate
trade-off to keep the platform simple for a small-scale study.

\subsection{Treatment Design: The Detection Warning}

The only difference between the two groups is a single informational
notice in the briefing and on the submission page for each task. The test
group sees the following text, displayed as a yellow banner on the task
screen:

\begin{quote}
\small
\textit{``Your final submission will be analyzed by an AI detection tool.
Your AI detection score will be recorded and included alongside your
submission in our research dataset.''}
\end{quote}

\begin{figure}[H]
    \centering
    \fbox{\includegraphics[width=0.95\textwidth]{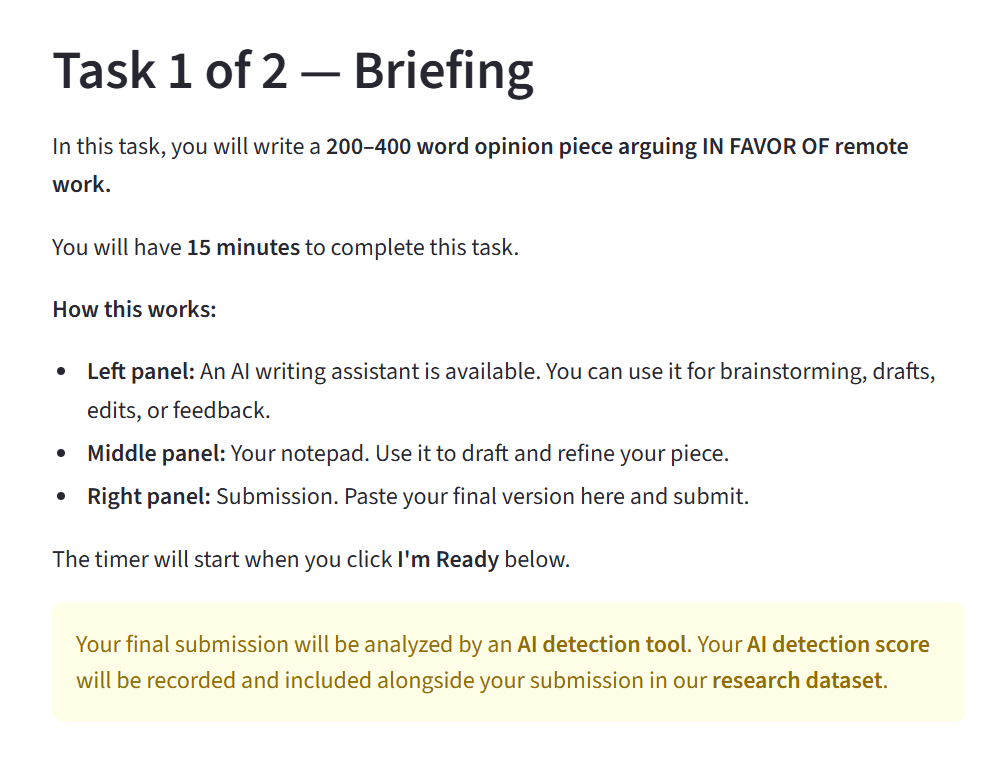}}
    \caption{The briefing for the test group. Notice the yellow tinted label for AI detection. Control \textit{does not} see this yellow label.}
    \label{fig:briefing_test}
\end{figure}

This warning was designed carefully. My overall goal was to reduce as
much behavioral and psychological bias as I could for the participants.
Research on ``priming'' distinguishes between warnings that tell people
what to do (behavioral primes, e.g., ``there is this feature in the
experiment and you should do this with it'') and warnings that activate
a concept without prescribing a response (conceptual primes, e.g.,
``there is an AI detector''). Conceptual primes can still change
behavior even when the prime itself does not specify an action
(Bargh et al., 1996), and a meta-analysis of incidentally presented
word primes confirmed that such effects are small but robust
(Weingarten et al., 2016). My warning
falls into that second category: it tells participants that detection
will happen, but it does not tell them what ``AI-sounding text'' looks
like or how to avoid it. The experiment does not even suggest there is
any consequence to being detected, nor that this is part of any real
evaluation. That ambiguity is the point.

I also had to be careful not to make the warning too aggressive. Research
on reactance shows that overly forceful warnings can backfire, causing
people to do the opposite of what you intended (Brehm, 1966). For this
experiment, I did not choose a large red banner with alarm language
because that could have pushed participants to use the AI with less or
more extreme intent (even going as far as defying the experiment), or
made them so anxious that their behavior would no longer reflect
anything natural. I chose a yellow background with neutral phrasing to
keep the warning noticeable but not alarming.

The label sits in 2 positions: one within the first briefing
just above the forced chatbot interaction to onboard the participant
(preventing cases where a participant might miss the label, Figure~1),
and the second moment is placed just above the ``I am ready to submit''
checkpoint to let the user see it again before submitting (Figure~3).

When designing the experiment, I considered how the user
might interact with the treatment itself when it comes to intent. The
treatment mirrors the kinds of warnings students might encounter in real
academic settings rather than creating an artificial challenge to beat
the detector. If the test group was motivated to defeat an AI detector,
it would lead to high-intent changes in behavior. I want to observe how people naturally responded to the idea of
being monitored, rather than feeling like they are playing a game.

\subsection{Demand Characteristics and the ``Observer Effect''}

In any experiment, there is a risk that participants figure out what you
are testing and change their behavior to match what they think you want.
This is called ``demand characteristics'' (Orne, 1962). There is also
another framework that separates participants being aware of the
hypothesis, being motivated to comply, and having a strategy for
responding. Awareness alone does not always lead to behavioral change
(Corneille \& Lush, 2023). A larger-scale test with over 12{,}000
participants found that explicitly telling people the hypothesis did not
\textit{actually} alter treatment effects, which suggests demand effects
may be less threatening than commonly assumed for this type of
experiment (Mummolo \& Peterson, 2019).

Still, I built in several protections to keep the test valid:

\begin{enumerate}[nosep]
    \item \textbf{Between-subjects design:} Each participant sees only one
          version of the experiment, so they have nothing to compare
          against.
    \item \textbf{Cover story:} The study is framed as research on
          ``how people use AI writing tools'' when recruiting volunteers
          (which is true but broader than the actual hypothesis).
    \item \textbf{Unobtrusive primary outcome:} The main thing I analyze
          is the submitted text itself, which participants produce
          naturally as part of the task.
    \item \textbf{Informational warning:} The treatment does not tell
          participants what the ``correct'' response is.
    \item \textbf{Post-experiment hypothesis probe:} After finishing, I ask
          ``What do you think this study was about?'' This lets me check
          whether anyone guessed the real purpose and filter accordingly.
          (No one has.)
\end{enumerate}

There is also a baseline observer effect to consider. Both groups know
they are in a study, which can change behavior on its own. This is a
common issue, however, a systematic review of 19 studies found that there
is no single observer effect (McCambridge et al., 2014). In this case,
both groups experienced the same baseline awareness of being observed.
The test group received a second layer of observation awareness (the AI
detection notice). So I am measuring the \textit{incremental} effect of
the detection warning above whatever baseline shift comes from being in
a study at all.

The closest parallel I found to this ``AI Detector'' label in practice is
plagiarism detection software like Turnitin. A study of medical students
found that introducing plagiarism detection software alongside penalties
reduced plagiarism rates from 21\% to 2\% across successive cohorts
(Bilic-Zulle et al., 2008). A similar
study at the University of Botswana found that warning students about
Turnitin reduced plagiarism from 20.5\% to 16.2\%, with the biggest
drops among the most egregious cases (Batane, 2010). But there is an
important difference between this study and the plagiarism research:
with plagiarism, students know exactly what to avoid (do not copy text).
In this study, participants are ambiguous about what the detector
targets. They do not know what ``AI-sounding text'' looks like. That
means the warning creates awareness without giving participants a clear
strategy for responding, which is what makes the behavioral response
interesting to study. The limitation of this study however is that it is inconsequential to the writers. There is no real
penalty for using AI here, unlike a school or a graded requirement.

\subsection{Writing Task Parameters}

Each participant writes two opinion pieces on remote work, one arguing
in favor and one arguing against, with a target length of 200 to 400
words per piece.

\textbf{Topic choice.} I chose remote work because it requires no
specialized knowledge, most people have an opinion on it, and it is
polarizing enough that writing against your own belief takes real
effort. That cognitive demand is intentional: it is exactly the
situation where reaching for the AI assistant becomes most tempting.

\textbf{Word count.} The 200 to 400 word target was intentionally set
based on a few pieces of evidence backed in prior research. Adults
composing original text produce roughly 19 words per minute (Karat et
al., 1999). With planning and pausing over 15 minutes, that works out
to roughly 200 to 350 words unassisted. Research on AI-assisted writing
found that AI tools reduce task completion time by about 40\% (Noy \&
Zhang, 2023), which in a fixed time window means roughly 280 to 490
words. On the analysis side, lexical diversity indices become
more stable as text length increases, with MTLD performing
reliably on texts as short as 50 words (Zenker \& Kyle, 2021),
so the 200-word floor comfortably supports the NLP features I
planned to extract without overwhelming participants. A 400-word ceiling
keeps the task achievable in 15 minutes but prevents participants from
just dumping raw AI output. The constraint forces them to at least
consider editorial choices about what to keep and what to cut, and that
curation is itself a signal worth measuring.

\textbf{Timer.} A soft 15-minute countdown runs on screen for each task.
It is not enforced: participants can continue writing after it reaches
zero. Participants are not aware of this soft countdown. The timer
creates psychological time pressure without risking data loss from a
hard cutoff. Actual duration is logged as a variable for analysis.

The other challenge is encouraging the use of the AI agent. Forcing
behavior with an AI agent is tedious to implement, and it would also
bias results by forcing individuals to write with an AI. Instead, the
briefing prior to the actual task forces a small demo / tutorial to
encourage the use of the AI assistant.

\subsection{Survey Design}

\textbf{Pre-survey.} Before writing, participants also report their name
or alias (anonymity is allowed), their personal stance on remote work,
their AI usage behaviors, and their education level.

The stance question uses a four-point scale with no neutral option:
Slightly Favor, Favor, Slightly Oppose, Oppose. The neutral option is
omitted on purpose. Research on survey design shows that including a
``no opinion'' or ``neutral'' midpoint encourages ``satisficing,'' where
respondents pick the middle option to avoid thinking about it (Krosnick et al.,
2002). Removing it forces everyone to lean one way, which gives a clean
way to classify each document as belief-aligned (the assigned stance
matches their real opinion) or belief-misaligned (it does not). That
classification is necessary for the belief-alignment analysis.

The AI usage question is multi-choice rather than a single frequency
measure. Participants check all that apply from a list that includes
``search engine use,'' ``writing help,'' ``coding and technical
tasks,'' ``work or school assignments,'' ``tried but irregular use,''
and ``does not use AI tools.'' This captures \textit{how} someone uses
AI, not just how often. A participant who uses AI for coding but never
for writing will likely interact with the writing assistant differently
than someone who uses AI for writing every day. For this sample size
that part is practically negligible, but I kept it for completeness.

\textbf{Post-survey.} After writing both tasks, participants self-reported
how much they relied on the AI for each piece using a five-point
ordinal scale ranging from ``I wrote it entirely myself'' to ``AI wrote
almost all of it''

The ordering of post-survey items mattered. Research on question order
effects shows that structured scales can frame how people respond to
open-ended questions if asked first (Schwarz, 1999). I put
open-ended questions before structured ones, and the hypothesis probe
(``What do you think this study was about?'') always comes last so it
cannot contaminate anything else.

\section{Platform, Implementation, and Data Collection}

\subsection{Platform Architecture}
\begin{figure}[p]
    \centering
    \includegraphics[width=\textwidth]{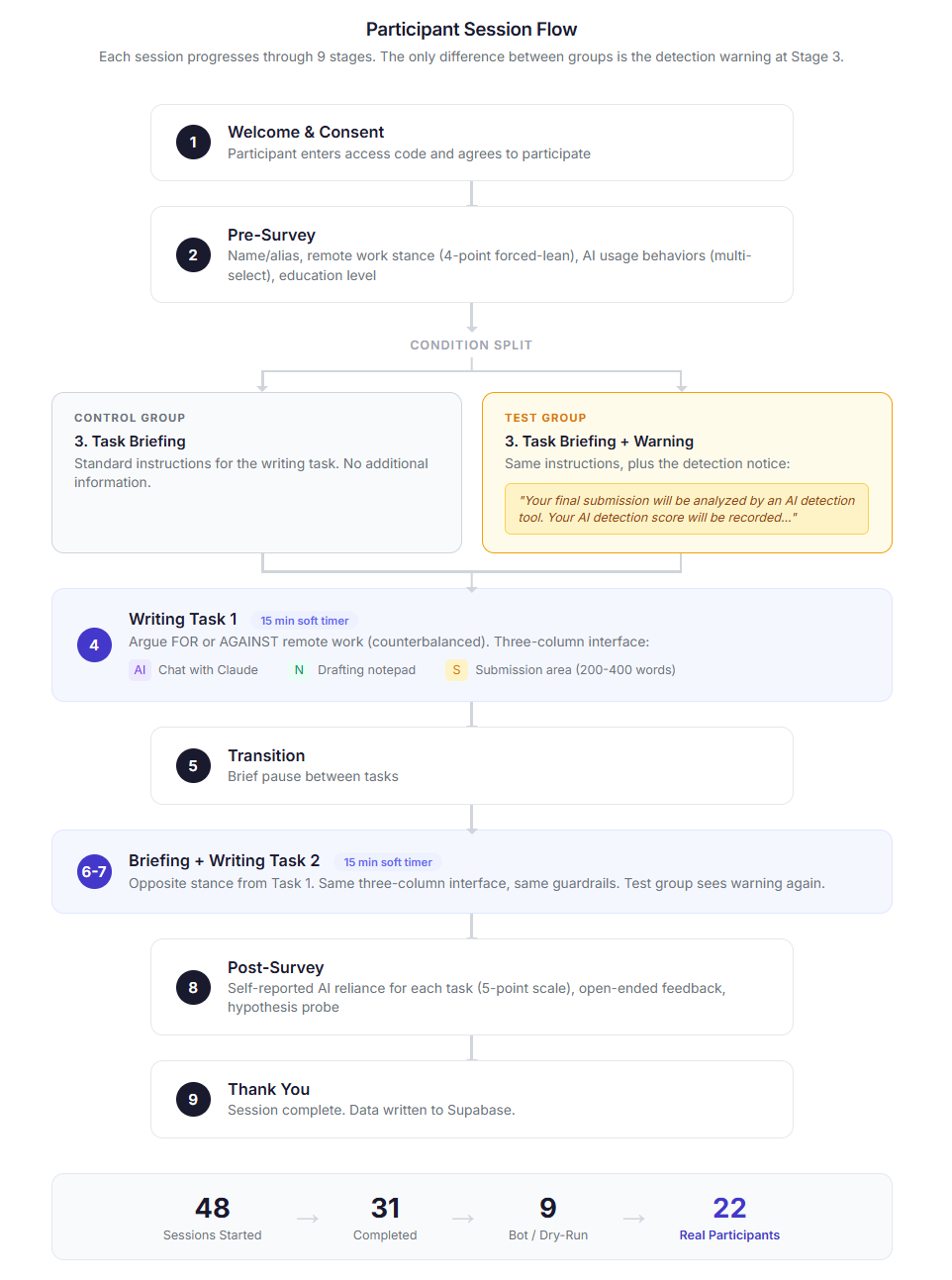}
    \caption{Participant session flow. Each session progresses through nine
    stages. The only difference between groups is the AI detection warning
    shown at Stage~3.}
    \label{fig:sessionflow}
\end{figure}
The experiment runs as a Streamlit app backed by the Anthropic API
(Sonnet was used as the base model) for the chatbot. All participants
access the same URL and enter a shared access code. The application
assigns conditions automatically based on the participant landing
sequence, manages the session flow, and writes data to a Supabase
PostgreSQL database in real time so I could analyze the results and
eventually propagate them into Phase~2 for the judges.

The session goes through nine stages in this static order:

\begin{enumerate}[nosep]
    \item Welcome and consent
    \item Pre-survey (demographics and stance)
    \item Briefing for Task~1 (test group sees the detection warning here)
    \item Task~1: writing with AI assistant
    \item Transition screen
    \item Briefing for Task~2
    \item Task~2: writing with AI assistant
    \item Post-survey (self-reported AI reliance)
    \item Thank-you screen
\end{enumerate}

The task screen uses a three-column layout. The left column contains
the AI chat interface, where participants can send messages to Claude
and receive responses in real time. The center column is a rich-text
notepad for drafting. The right column holds the submission text area
with a live word count and a submit button. This separation between
drafting space and submission space forces participants to make a
deliberate editorial choice about what to submit rather than pasting
raw AI output directly.

\begin{figure}[H]
    \centering
    \fbox{\includegraphics[width=0.95\textwidth]{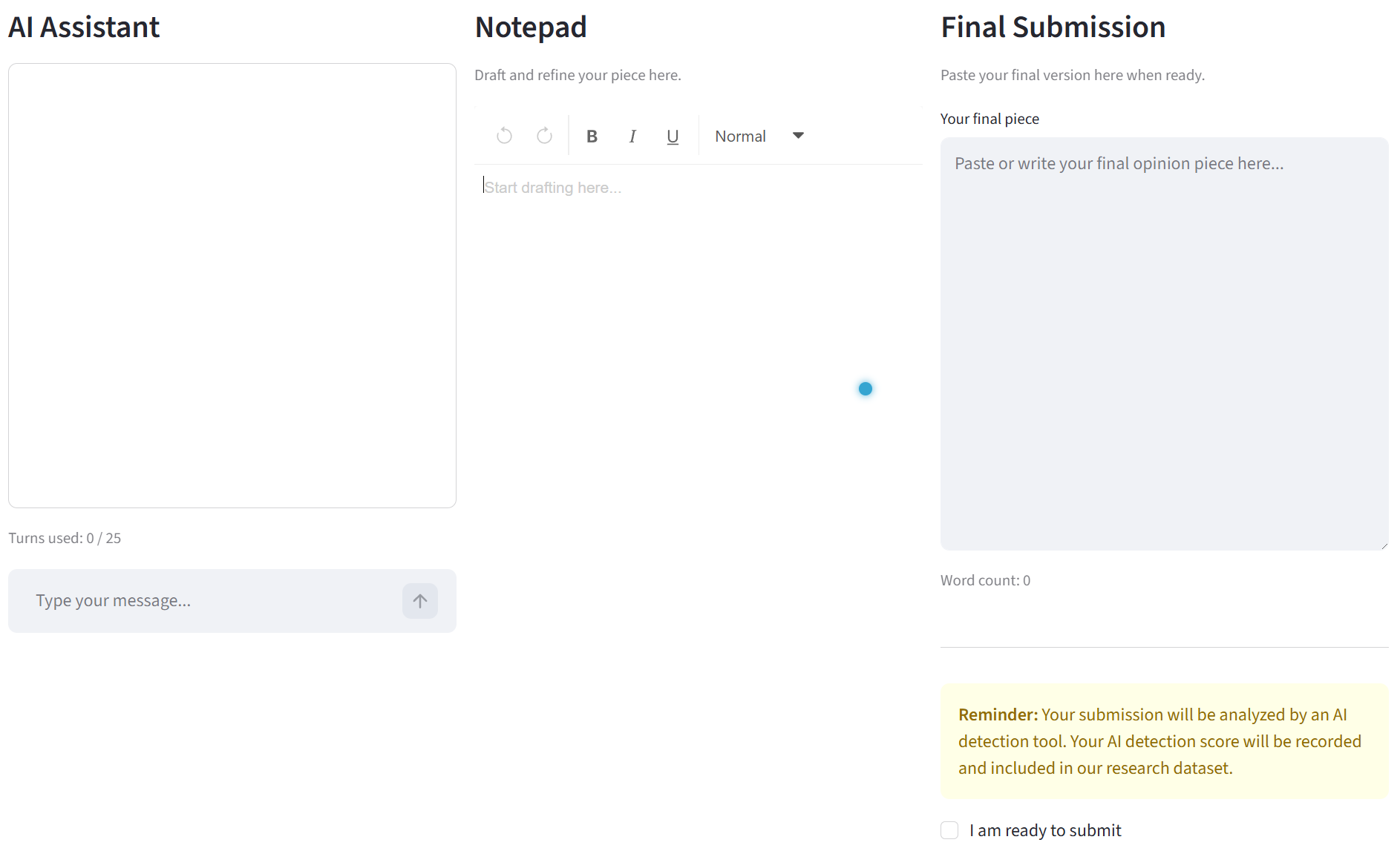}}
    \caption{Both groups see the same three-panel setup, with an AI chatbot, a notepad for edits, and the final submission screen. Users confirm with the button if they are ready to submit to avoid accidents. Only the test group sees the AI detection label.}
    \label{fig:task_ui}
\end{figure}

\subsection{Recruitment and Data Collection}

Participants were recruited through personal outreach via social media
(Instagram, LinkedIn) and direct messaging (WhatsApp), targeting friends,
classmates, and broader social networks. The response rate among even
close acquaintances was modest---a reminder that the goodwill of friends,
while freely given in conversation, becomes a scarcer resource when a
hyperlink and twenty minutes are involved.

Data is stored in Supabase, a hosted PostgreSQL service. Each session
is a single row containing the full interaction record: pre-survey
responses, conversation logs with timestamps and token counts for every
message, submission text and word count, post-survey responses, and
session-level metadata (start time, completion time, total tokens
used).

\subsection{Completion Funnel}

\textbf{Phase 1.} Of the sessions initiated during the collection
period, a subset reached the ``thank-you'' screen. After filtering out
bot dry-runs and incomplete sessions, the final analytic sample is
\textbf{21 real Phase~1 participants} producing \textbf{41 validated
documents} (one participant had a single completed task, the rest
contributed both). The dropped sessions were participants who entered
the access code but did not finish, typically due to time, loss of
interest, or mobile-browser issues.

\textbf{Phase 2.} For the judge evaluation, \textbf{315 judge sessions
were started}, \textbf{76 sessions completed fully}, and together they
produced roughly \textbf{2{,}000 pair-level responses} (partial
completions included, since each individual pair response is still
usable data)

\section{Phase 1 Results (RQ1)}

This section covers the EDA of the Phase~1 participants and the
41 validated submissions. Its purpose is to answer research question 1.
Do the two groups within the study
produce different writing patterns and writing outputs? With 9 control and 12 treatment
participants completing the study, the sample is underpowered to run any statistical tests on the differences between both groups. However, there are still basic descriptive patterns that show both groups treated the study differently.

\subsection{Sample Demographics}

The sample is skewed toward writers with a graduate-level education, reflecting the
Georgia Tech and personal-network recruitment pool from social media and messaging forums. The "Remote-work" stance
is heavily favorable: most completed participants selected ``Favor'' or
``Slightly Favor,'' with only a small minority selecting the opposing
side. The practical consequence is that most ``Against'' documents were
written in a belief-misaligned condition, which the judges might mark differently. Condition assignment produced 9 control and 12
treatment participants (18 and 23 validated documents respectively since participants complete two documents).
The imbalance is the result of session drop-off or other incomplete submissions being removed, not of the
assignment rule itself. Section 2.1 describes how a deterministic rule in the Streamlit app would ideally produce an exact 50/50 split on any complete sample.

\subsection{Chatbot Interaction by Condition}

\begin{figure}[H]
\centering
\includegraphics[width=0.92\textwidth]{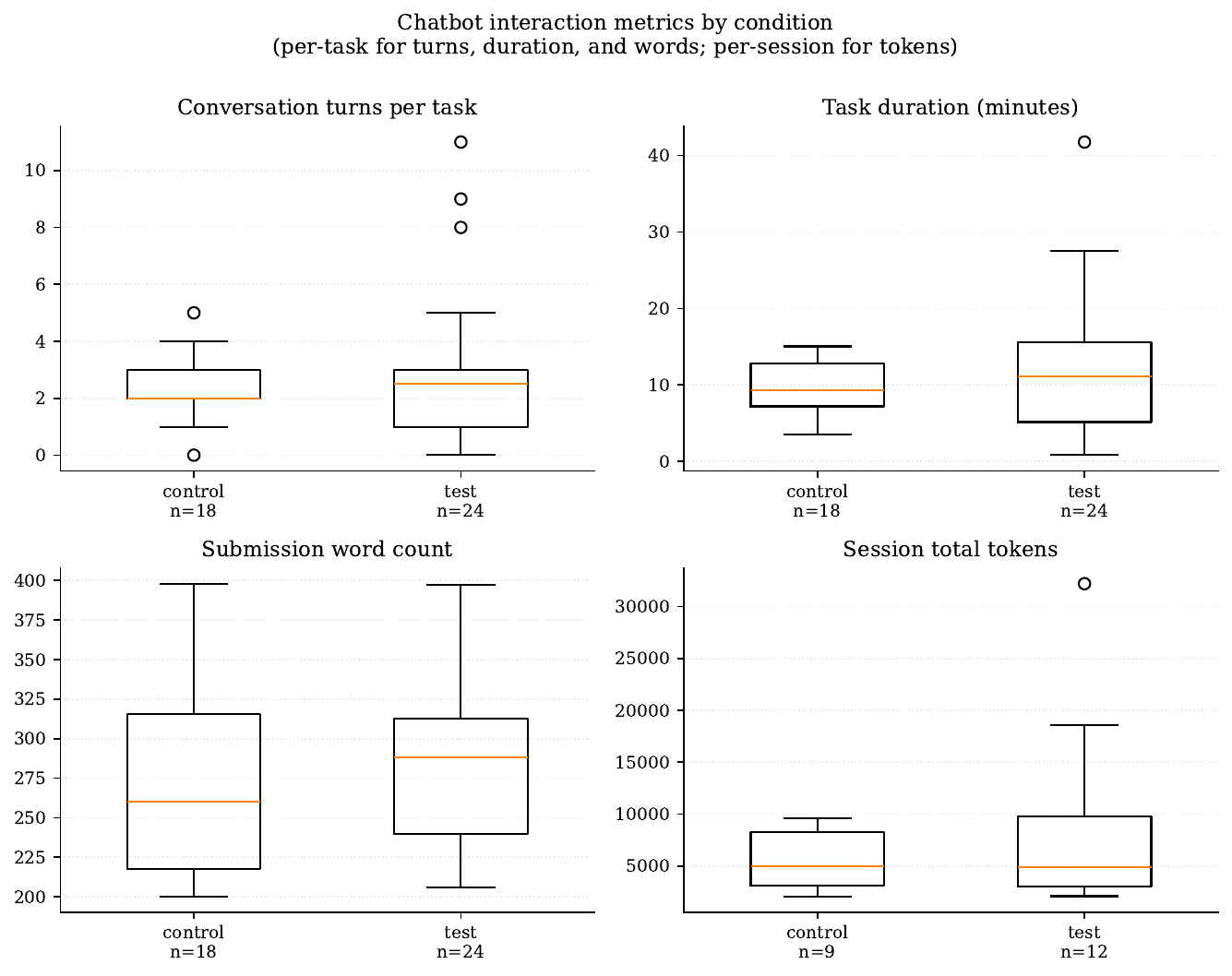}
\caption{Chatbot interaction metrics by condition. Turns, duration, and
submission word count are reported per task (two tasks per participant).
Session total tokens are reported per participant. Per-task counts
reflect all completed session tasks before document-level validation
filtering (one test-group task was excluded during validation,
reducing the validated document count to 23).}
\label{fig:process-by-cond}
\end{figure}

Table~\ref{tab:behavioral-final} summarizes the main chatbot-interaction
metrics per task, split by condition, and Figure~\ref{fig:process-by-cond}
shows the same data as box plots.

\begin{table}[H]
\centering
\small
\caption{Chatbot interaction metrics by condition. Turns, duration, and
word count are per task; session total tokens are per participant.}
\label{tab:behavioral-final}
\begin{tabular}{@{} l r r r r @{}}
\toprule
 & \multicolumn{2}{c}{\textbf{Control ($n=9$)}}
 & \multicolumn{2}{c}{\textbf{Treatment ($n=12$)}} \\
\cmidrule(lr){2-3}\cmidrule(lr){4-5}
\textbf{Metric} & \textbf{Mean} & \textbf{Median}
                & \textbf{Mean} & \textbf{Median} \\
\midrule
Conversation turns (per task)   & 2.4  & 2.0  & 3.0  & 2.5 \\
Task duration (seconds)         & 571  & 560  & 693  & 668 \\
Submission word count           & 275  & 260  & 278  & 288 \\
Session total tokens            & 5{,}411 & 5{,}010 & 8{,}717 & 4{,}876 \\
\bottomrule
\end{tabular}
\end{table}

Three descriptive patterns are visible. First, treatment participants
take more conversation turns per task than control participants on
both the mean and the median. Second, treatment participants generally spend
longer on each task by about two minutes at the median.
Third, session total tokens are higher on average for the treatment
group but with much wider variance, driven by a few participants who
had long back-and-forth sessions with the chatbot. Word
counts are nearly identical across both groups, which is expected
given that the task prompt bounded the length target at 200 to 400
words. Qualitatively, some users simply prompted the chatbot to give it an essay spanning the word count. These patterns answer RQ1 at the descriptive level: the two
groups do interact with the chatbot differently, with the treatment
group using it more intensively.

\subsection{AI Overlap with Chatbot Output}

To quantify how much of each submission came from the chatbot, I
feature engineered an AI overlap score comparing the submitted text to the
assistant messages in that participant's conversation log.

\paragraph{Motivation and design rationale.} The simplest approach to
``how much of the submission came from the chatbot?'' is exact
word-for-word matching. I avoided using this metric alone because this doesn't capture edge cases in two distinct ways.
First, a writer can keep the AI's structure and substitute
synonyms or different phrases. This won't be captured. A writer can also try reordering text, and this only counts the
sentences that survived in order. A "score" for this setting
therefore needs to combine multiple similarity metrics rather than rely
on one. Document fingerprinting techniques use n-gram hashing
to detect reused passages across texts (Schleimer et al., 2003),
and large-language-model training-data decontamination uses
similar n-gram overlap and exact substring matching via suffix
arrays (Brown et al., 2020; Lee et al., 2022). I adapted the
general principle of combining multiple text-similarity measures
to engineer a robust scoring function.

\paragraph{Definition.} Let's make $S$ a participant's submission text,
(lowercased and with whitespace normalized), and let
$A_1, A_2, \ldots, A_k$ be the assistant messages (or turns) in that
participant's conversation log for the same task, preprocessed the
same way. For any single assistant message "$A$", I compute three
similarity scores between $S$ and $A$. Each one targets a different
way a writer might reuse AI text. This is to capture the AI overlap the user submitted.

The first score looks at short phrases. We can use a type of sliding-window method to capture similarity. For example, Let $T_3(x)$ be the set of
three-word windows in $x$ (for example, the sentence ``the dog sat
down'' contains the windows ``the dog sat'' and ``dog sat down''). I
measure what fraction of the submission's three-word windows also
appear somewhere within the AI messages sent to the user. This is very similar to a Jaccard Score but using full-words instead of n-grams:
\begin{equation}
t(S, A) \ = \ \frac{|\,T_3(S) \,\cap\, T_3(A)\,|}{|\,T_3(S)\,|}.
\end{equation}
A high value here flags short phrase-level reuse, even when the
writer placed fragments from different AI responses together.

The second score looks for one long paste. I find the
longest run of characters that appears identically in both $S$ and
$A$, call its length $L(S, A)$, and divide by the length of the
submission. This is identical to looking at a word-for-word overlap assuming they just copy-pasted and didn't make any edits:
\begin{equation}
\ell(S, A) \ = \ \frac{L(S, A)}{|\,S\,|}.
\end{equation}
A value near 1 means almost the entire submission is a single copy of AI text. The same exact function is used in plagiarism detection
(Schleimer et al., 2003) and in large-language-model training-data
decontamination (Brown et al., 2020; Lee et al., 2022).

The third score captures total overlap, including pieces that have
been shuffled around. Using a standard sequence-matching routine
from the Python standard library, I find every non-overlapping run
of characters that appears in both $S$ and $A$ and sum their
lengths, giving $M(S, A)$.
\begin{equation}
r(S, A) \ = \ \frac{2 \, M(S, A)}{|\,S\,| \,+\, |\,A\,|}.
\end{equation}
This score would stay closer to 1 when a writer kept most of the AI's sentences or words
but rearranged sentences or paragraphs. The previous two
measures would each miss on their own.

To summarize how much $S$ resembles a given AI message $A_i$, I
take the largest of the three scores:
\begin{equation}
AI~overlap~Score\ = \ \max\bigl(\,t(S, A_i),\ \ell(S, A_i),\ r(S, A_i)\,\bigr).
\end{equation}
The AI overlap score for the document is then the largest of the set of scores that we calculate.
A score near 1 means the submission is exactly a paste of one of the
AI responses the writers had with the chat. A score near 0 means the submission shares no
meaningful text with anything the AI produced, or the participant
never actually used the chatbot.

\paragraph{Why three measures and not one?} Each of the three
checks a different way a writer might reuse AI text:
small phrases, long chunks, or the full AI output reordered in some way.
Taking the maximum is the conservative choice. The idea is that
a writer cannot circumvent the scoring if they utilized the AI by rearranging or lightly paraphrasing. However, there may be some edge cases or limitations to these methods, but I used them to cover the main use-cases of using AI-assistance for writing.

\begin{figure}[H]
\centering
\includegraphics[width=0.65\textwidth]{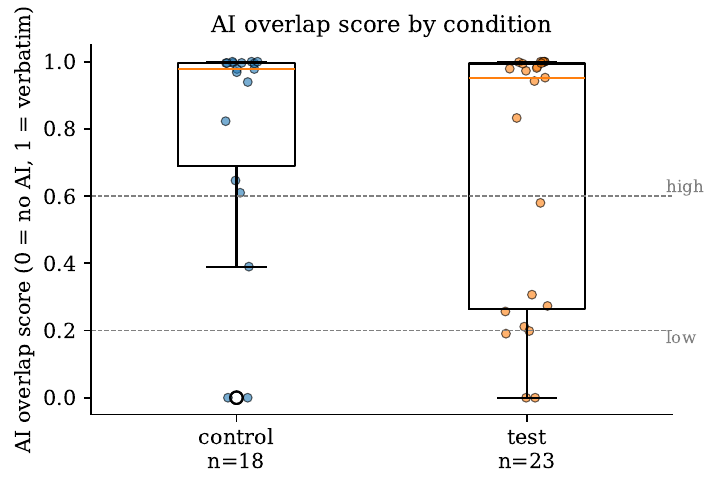}
\caption{AI overlap score by condition. The score is the maximum
similarity between the submission and any single AI response across
three measures. Note how the test group has more cases where users used AI infrequently or not at all for their writing}
\label{fig:overlap-by-cond}
\end{figure}

\begin{figure}[H]
\centering
\includegraphics[width=0.7\textwidth]{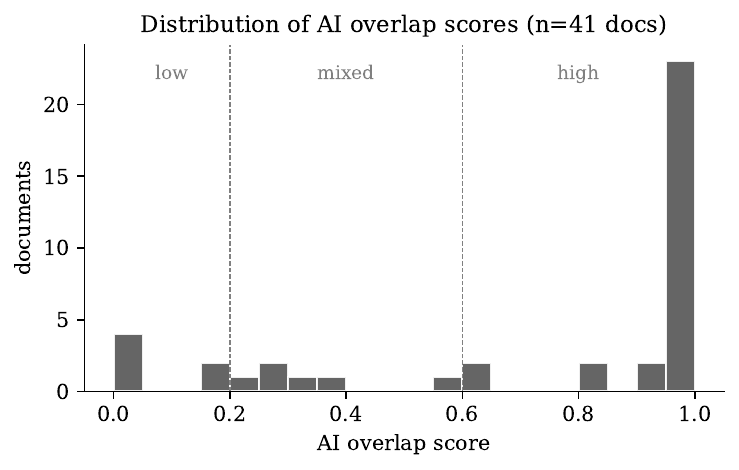}
\caption{Distribution of AI overlap scores across all 41 validated
documents. The distribution for the writers is concentrated on many writers who used AI verbatim in some way according to the three measures, while some documents used more editing and a handful used none.}
\label{fig:overlap-dist}
\end{figure}

\paragraph{What the score shows.} Figure~\ref{fig:overlap-by-cond} shows
the overlap-score distribution by condition.
Figure~\ref{fig:overlap-dist} shows the pooled histogram across all 41
documents.

Only six documents sit at or near zero. Of these, three
participants never sent a user message to the chatbot on one of their
tasks, one participant sent a message but the session recorded no
assistant response (this was a streamlit bug), and two participants iterated with the chatbot but
retained little of its output in the submission. The remaining 35
documents concentrate above 0.6, with a mode near 1.0. The control and
treatment medians are 0.978 and 0.953 respectively, nearly identical.
At the descriptive level, most writers in both conditions pasted
at least one AI response close to verbatim, and a minority in each
condition either did not engage the chatbot at all or rewrote
substantially.

\subsection{Stylistic (Stylometric) Features}

\begin{figure}[H]
\centering
\includegraphics[width=0.85\textwidth]{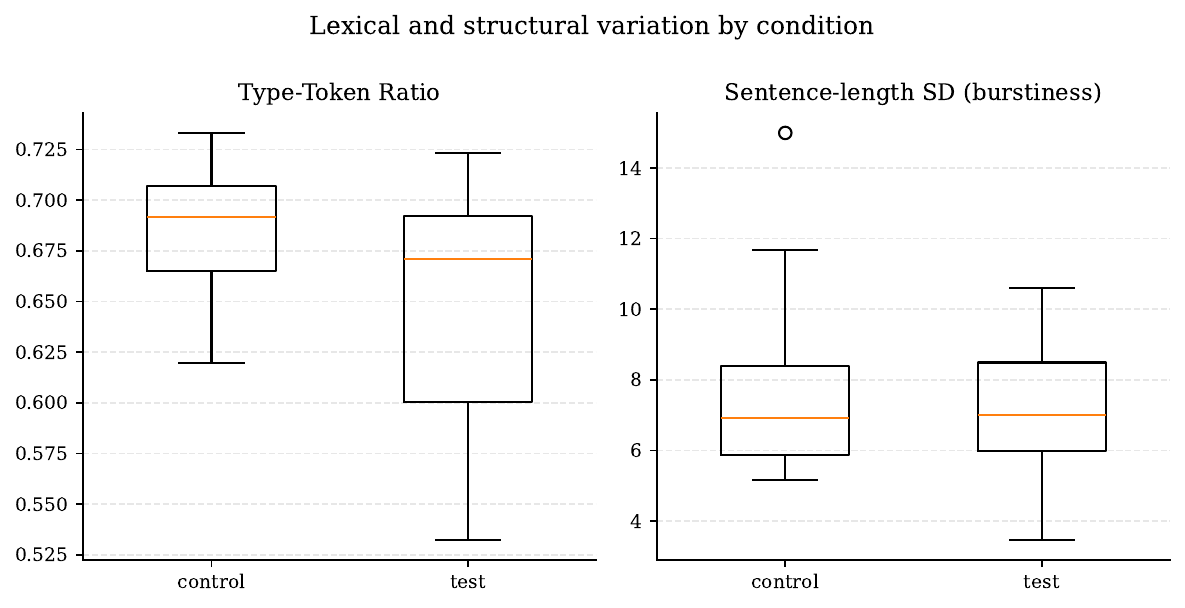}
\caption{Lexical and structural variation by condition: type-token
ratio and sentence-length SD (burstiness). Distributions overlap
heavily between control and treatment.}
\label{fig:features-by-cond-a}
\end{figure}

\begin{figure}[H]
\centering
\includegraphics[width=0.85\textwidth]{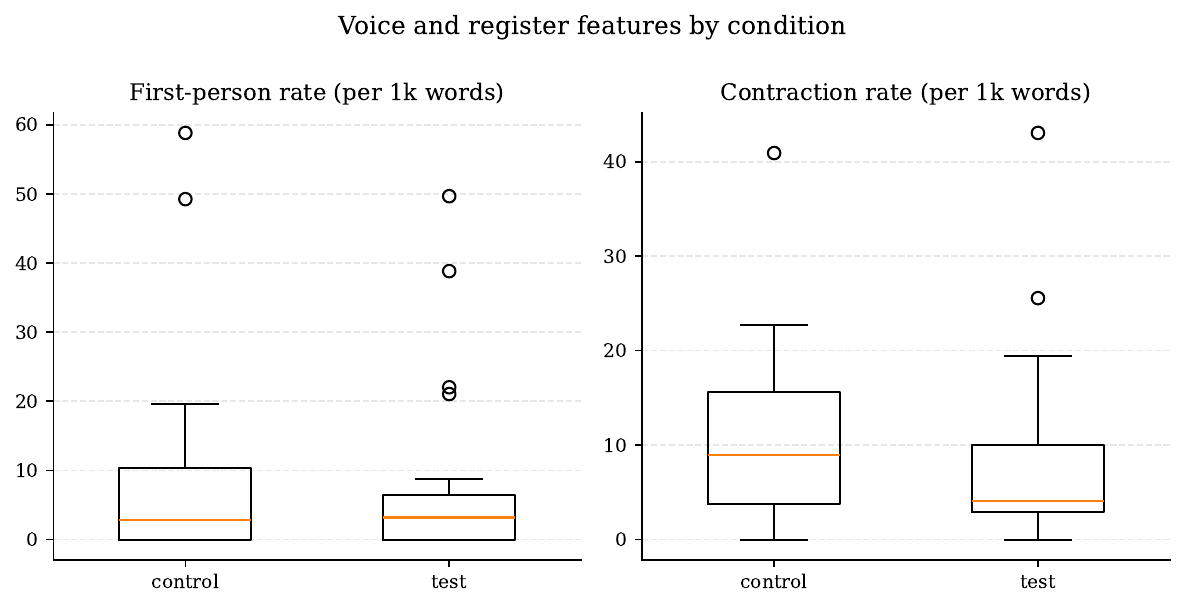}
\caption{Voice and register features by condition: first-person rate
and contraction rate (both per 1{,}000 words). Distributions overlap
heavily between control and treatment.}
\label{fig:features-by-cond-b}
\end{figure}

Table~\ref{tab:stylometric-features} lists the stylometric features
extracted from each submission. Each feature has an established role
in stylometry or AI-text detection and is cheap to compute.
Figures~\ref{fig:features-by-cond-a} and~\ref{fig:features-by-cond-b}
show four of them as box plots by condition.

\begin{table}[H]
\centering
\small
\renewcommand{\arraystretch}{1.25}
\caption{Stylometric features computed per submission. $N$ is the
number of words and $N_s$ the number of sentences. ``per 1k'' rates
multiply the token count by $1000/N$.}
\label{tab:stylometric-features}
\begin{tabular}{@{} p{3.3cm} p{5.3cm} p{5.7cm} @{}}
\toprule
\textbf{Feature} & \textbf{Definition} & \textbf{What it captures / source} \\
\midrule
Type-Token Ratio
& $\mathrm{TTR} = |V| / N$, where $V$ is the set of unique tokens
& Lexical diversity. AI output tends to reuse the same words more often within a response (Reviriego et al., 2024). \\
\addlinespace[0.5em]
Mean sentence length
& $\bar{L} = N / N_s$
& Baseline rhythm. AI output tends toward longer, more uniformly structured sentences. \\
\addlinespace[0.5em]
Sentence-length SD
& $\sigma_L$ across sentences
& Variation in rhythm. Human writers mix short and long sentences more often; AI output tends to be more uniform (Desaire et al., 2023). \\
\addlinespace[0.5em]
First-person rate (per 1k)
& $\mathrm{count}(\text{I, me, my, we, our, \ldots}) \cdot 1000 / N$
& Personal voice. AI output typically uses fewer first-person pronouns than human writing (Sandler et al., 2024). \\
\addlinespace[0.5em]
Hedging rate (per 1k)
& $\mathrm{count}(\text{perhaps, might, possibly, \ldots}) \cdot 1000 / N$
& Tentative language. Epistemic stance, as reflected in lexical bundles, differs in frequency between human and AI text (Jiang \& Hyland, 2025). \\
\addlinespace[0.5em]
Contraction rate (per 1k)
& $\mathrm{count}(\text{don't, can't, it's, won't, \ldots}) \cdot 1000 / N$\footnotemark
& Casual tone. AI output often avoids using contractions, preferring formal phrasing like ``do not'' instead of ``don't.'' \\
\bottomrule
\end{tabular}
\end{table}
\footnotetext{The contraction regex also matches possessive forms (e.g., ``student's''). Both groups are affected equally, so this does not bias the comparison, but the rate slightly overstates true contraction frequency.}

\subsection{Summary of Phase 1 EDA}
My research question 1 asked whether writers behave differently under the threat of AI detection.
With such a small sample size of writings, we can't run any statistical tests to genuinely prove the treatment group behaved differently in the study compared to the control group. However, based on the EDA and the analysis of different features between both groups, there are differences in behaviors between both. This is where evaluation from the judges for these documents becomes critical to the success of this experiment. The judges \textit{become} the statistical test of this study - where they evaluate whether or not the behavior of the treatment group would create more "human-like" submissions. The experiment design almost intends to compensate for the lack of power in a statistical test between both groups - and the judges become a secondary layer from a human perspective to determine whether a behavior difference occurred when the treatment group saw the AI-detection label. In phase 1, I found that the
treatment group takes more conversation turns, spends more time on
each task, and exchanges more tokens per session than the control
group. But the submissions completed tell us that their usage of AI text is almost identical. Control and
treatment AI similarity scores differ by only 0.025 between the medians, and the stylometric
feature distributions overlap as well. Both groups relied heavily on the AI chatbot to complete the writing within the word-count and the time alloted, but the treatment group was shown to be more variable in their interactions with the AI - some completely abandoning the chatbot, or some iterating extensively to deliver a submission. Phase~2 becomes the statistical test for the difference between these two groups. Where judges will compare both documents under a deceptive design and asked "Which document was written by a human" - despite both being written by AI. The hypothesis is that the group seeing the AI-detection threat will be interpreted as more likely to be "human" written compared to the control group's submissions.

\section{Phase 2: Human Judge Evaluation (RQ2)}

\subsection{Recruitment and Completion}

\begin{figure}[H]
\centering
\includegraphics[width=.9\textwidth]{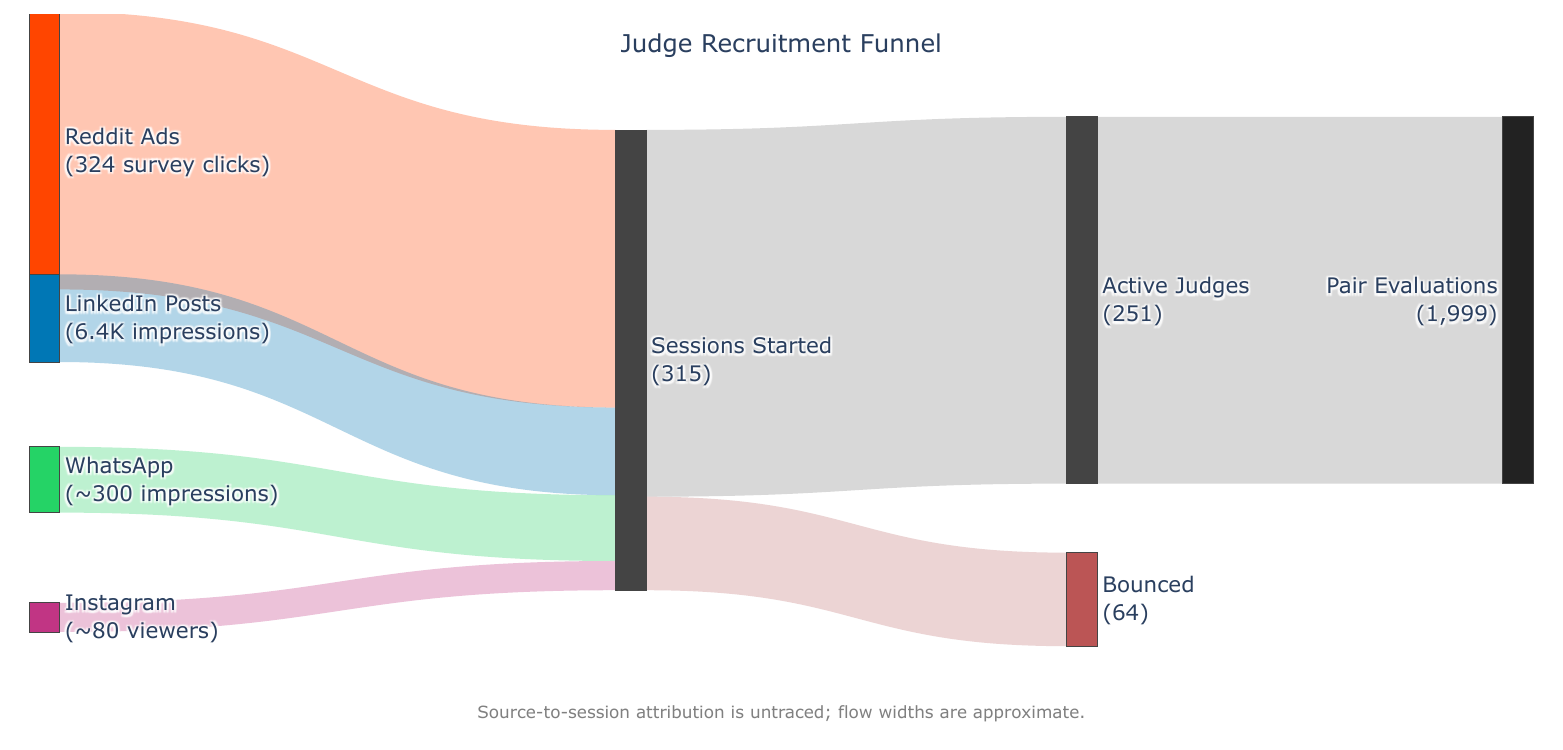}
\caption{Judge recruitment funnel. Session attribution is
untraced; flow widths from the four channels are approximate.}
\label{fig:recruitment-sankey}
\end{figure}

I recruited judges through four channels over a three-week period:
Reddit ads (~32,000 impressions, 1,200 post landings, 324 survey clicks),
LinkedIn posts across four rounds (~6,400 total impressions),
direct WhatsApp messages to roughly 300 contacts, and Instagram posts
(approximately 80 viewers). Attribution is untraced
because the survey link was identical across all channels, but Reddit
was the dominant source by volume.
Figure~\ref{fig:recruitment-sankey} shows the recruitment funnel.

Of 315 sessions started, 251 judges submitted at least one
evaluation, producing 1{,}999 total pair-level responses. Each pair
showed one control document and one treatment document for the same
stance (both ``For'' or both ``Against'' remote work), with the left/right
position randomized. The judges saw a 150-word preview of each document with a forced 10-second reveal period before they could choose, acting as a guard-rail against automatic responses and scripting. A judge could also optionally expand to read the full text. The survey asked
one question per pair: ``Which one was written by a human?'' with an optional confidence rating.

\subsection{Detection Accuracy}

Across all 1,999 paired evaluations, judges selected the treatment
(warned) document as the human-written one 54.13\% of the time,
versus 45.87\% for the control (unwarned) document. A two-sided binomial test against the chance rate of 50\%, treating each
response as independent, rejects the null hypothesis that this is a random guess at
$p = 0.000243$ ($\alpha = 0.05$). The 95\% Wilson confidence interval
on the control-as-human rate is [43.70\%, 48.06\%], which sits
entirely below the 50\% chance line. The effect seen is consistent across
both stances. FOR-stance pairs show a 46.00\% control-as-human
rate ($p = 0.012$, $n = 1{,}000$), and AGAINST-stance pairs show
45.75\% ($p = 0.008$, $n = 999$).

\begin{figure}[H]
\centering
\includegraphics[width=1\textwidth]{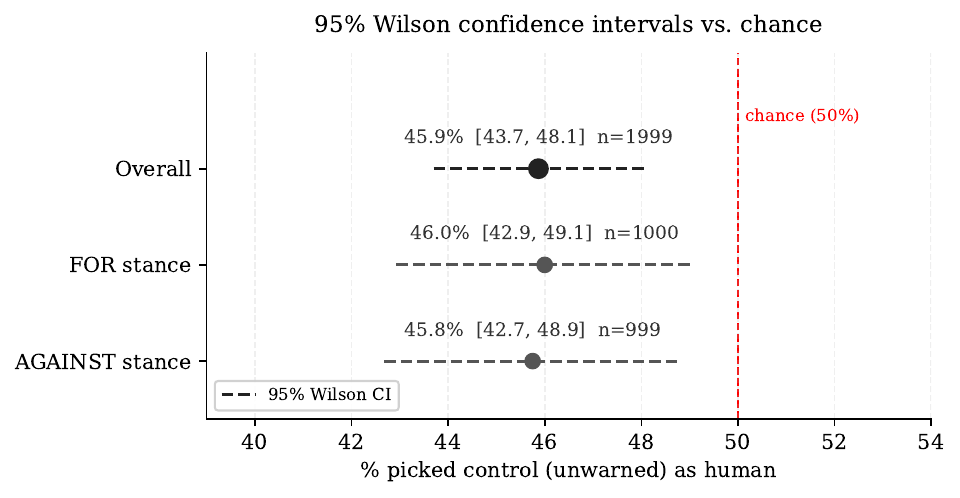}
\caption{95\% Wilson confidence intervals for the rate at which judges
picked the control (unwarned) document as human. The dashed red line
marks the 50\% chance rate. All intervals fall below 50\%, consistent
with the hypothesis that warned writers are perceived as more human.}
\label{fig:judge-ci}
\end{figure}

Figure~\ref{fig:judge-ci} shows these intervals. All
three (overall, FOR, AGAINST) fall to the left of
the 50\% chance line, showing that the effect is not driven by one
stance and is statistically significant.

The hypothesis was that writers in the treatment group who saw the AI-detection warning
would edit their submissions more carefully to sound human, and that judges would
therefore identify the warned (treatment) document as human at a higher
rate. The observed effect is in exactly that direction. The magnitude
is small, but it is statistically
significant across both stances.

\subsection{Confidence, Reading Behavior, and Fatigue}

\begin{figure}[H]
\centering
\includegraphics[width=0.65\textwidth]{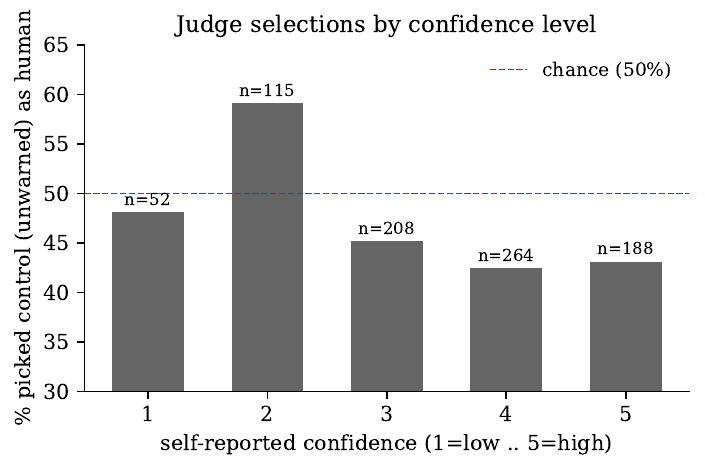}
\caption{Control-as-human rate by judge self-reported confidence
(1 = guess, 5 = confident). Higher-confidence judges leaned more toward
selecting the treatment document as human.}
\label{fig:judge-confidence}
\end{figure}

\paragraph{Confidence.} After each pair, judges had the option to rate their confidence
on a scale from 1 to 5 (1 being a guess, 5 being very confident). Higher-confidence judges were more likely
to pick the treatment document as human At
confidence levels 4 and 5, the control-as-human rate drops to
42 to 43\%, compared to the overall 45.87\%.
Figure~\ref{fig:judge-confidence} shows the breakdown by confidence
level.

\begin{figure}[H]
\centering
\includegraphics[width=0.65\textwidth]{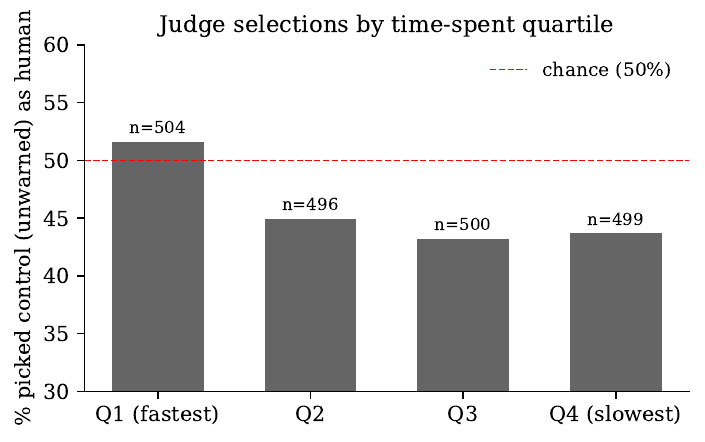}
\caption{Control-as-human rate by time-spent quartile. The fastest
responders (Q1, under 10.6s) are near chance; slower readers lean
toward selecting the treatment document as human.}
\label{fig:judge-time}
\end{figure}

\paragraph{Reading time.} Judges who spent more time on a pair were more likely to pick the treatment document as human. The
fastest quartile (under 10.6 seconds) was the closest to being near chance at 51.6\%, which is intuitive when judges were only briefly glancing at the writings rather than reading thoroughly.
The three slower quartiles ranged from 43\%
to 45\%.
Figure~\ref{fig:judge-time} shows the breakdown by time quartile.

\paragraph{Document expansion.} Most responses (76.3\%) involved no
expansion of the document at all. This meant judges decided from the 150-word
preview alone in the majority of cases. Among the 92 responses where the judge expanded \textit{just} one of the two documents, the control-as-human rate dropped to 35.9\%. Meaning that when a judge expanded a writing to investigate, they were able to pick the treatment document as human 64.1\% of the
time. What is interesting is that when judges expanded both documents - this brought the rate back to 45.8\%. This suggests that selectively reading
one document in full may amplify whatever cue judges are picking up on. However, given the small sample of judges who deliberately expanded documents, it is possible that this finding is simply noise - a judge may have expanded documents just to check if the UI worked rather than actually reading the expanded text.

\paragraph{Fatigue over time}
I split each judge's responses into terciles (early, middle, late)
to check whether detection accuracy changed over the course of a
session. The control-as-human rate was 45.87\% for early pairs,
45.74\% for middle, and 46.03\% for late. There is no clear fatigue
effect present. Judges didn't get better or worse over time, and the signal
held steady throughout each session.\footnote{This analysis was motivated by
qualitative feedback from a Reddit commenter during judge recruitment
who suggested that detection ability might degrade as judges reviewed
more pairs.}

\subsection{Per-Document Analysis: Is the Effect Driven by Outliers?}

I checked the concern that it was possible within a small document pool that one or two
outlier documents are driving entire experimental result. To check, I computed
the rate at which each of the 41 documents were chosen as human across
all the pairs it appeared in.

\begin{table}[H]
\centering
\small
\renewcommand{\arraystretch}{1.15}
\caption{Per-document chosen-as-human rates, summarized by condition.
The effect is distributed: most treatment documents sit above 50\%
and most control documents sit below it.}
\label{tab:per-doc-rates}
\begin{tabular}{@{} l c c c c @{}}
\toprule
\textbf{Condition} & \textbf{Docs} & \textbf{Mean rate}
  & \textbf{Docs above 50\%} & \textbf{Range} \\
\midrule
Control  & 18 & 45.9\% & 5 of 18 (28\%) & 29.8\% to 62.4\% \\
\addlinespace[0.3em]
Treatment & 23 & 54.0\% & 15 of 23 (65\%) & 30.5\% to 76.5\% \\
\bottomrule
\end{tabular}
\end{table}

The strongest outlier is participant P26 (treatment), whose two
documents were chosen as human 72\% and 76\% of the time. However, this makes sense entirely. P26 \textit{never}
sent a single message to the AI chatbot. They wrote both submissions
entirely on their own, and their text reads accordingly. They have frequent
contractions, personal phrasing, and variable sentence lengths. Judges may have recognized this as distinctly human. Removing P26 from the dataset
weakens the overall effect (control-as-human rate moves from 45.87\%
to 47.89\%) but \textit{does not} eliminate it, and the broader pattern holds:
15 of 23 treatment documents were chosen as human above the 50\%
chance line, compared to only 5 of 18 control documents. The effect
is distributed across the document pool, not concentrated in a single
outlier.

\subsection{Summary of Phase 2}

Phase~2 confirmed the original hypothesis: judges identified the
warned writer's document as human at a rate above chance. The effect
is small but statistically significant and holds across both stances,
across confidence levels, and across the full session without fatigue.
No single document drives the result.

\section{Discussion and Conclusion}

We reject the null hypothesis that there is no difference between the treatment (warning) group and the control (no warning) group. Judges identified the treatment document as "human" at a statistically significant rate more often than they identified the control document as being human. This is consistent with the original hypothesis for this experiment. The effect is
small but significant and holds across both stances - both FOR and AGAINST.

What is interesting to note is given a warning of AI Detection, participants \textit{did not} necessarily shy away from using an AI chatbot to assist them in their writing. Instead, the \textit{opposite effect} was seen.
Writers spent more time, took more turns, and used more tokens than the control group. Several treatment participants abandoned using the chatbot entirely, choosing to write the whole submission themselves. No control group participant did this. The warning \textit{didn't} mitigate the use of AI; it actually polarized writers to lean harder into using AI or opting out completely.

Yet, for all that behavioral difference between the two groups, the submissions themselves were nearly identical when measured for AI similarity to what was used in the chats. AI scores only differed by 0.025 between medians. Every stylistic feature showed significant overlap between both groups. The judges however chose the treatment document as "human" more often, at a statistically significant rate across ~2000 paired evaluations.

The implication here tells us that warned writers difference in behavior - using AI or not - changed the way their submissions were interpreted by human judges. This could mean a stylistic rhythm, or a personal opening, or a different prose within their conversations with the AI chatbot or their own drafts, is not a measurable feature analytically - but a human judge could identify the difference themselves.

Feature-based AI detection tools may extract typical AI signals, but this experiment suggests that humans given an AI-detection flag can actually lead to submissions that can convince judges to choose these submissions as more 'human' sounding.

\subsection{Limitations}

Several limitations constrain generalizability. The sample of 21 writers is small, skewed toward graduate students, and recruited from personal networks. The 9-vs-12 condition split limits statistical power for within-Phase~1 comparisons. The writing task (opinion pieces on remote work) may not generalize to other genres or domains. The AI overlap score captures surface-level textual reuse but not deeper forms of AI reliance such as idea generation or structural borrowing. The warning carries no real consequence, unlike academic integrity policies; a consequential warning might produce stronger effects. The judge task is inherently noisy---while the aggregate effect is significant, individual judge accuracy is only slightly above chance.

We assume each observation and judgment of the sample pairs is independent from another, but responses are clustered by both judge (251 judges contributing ${\sim}$8 responses each) and document (41 documents each appearing in many pairs). The effective sample size is therefore smaller than the 1{,}999 responses reported, and the binomial p-value ($p = 0.000243$) may be anti-conservative. A mixed-effects logistic regression with crossed random intercepts for judge and document would provide a more conservative test. We note, however, that the effect is consistent across multiple independent decompositions---by stance, by confidence level, by time quartile, and per-document---which provides converging evidence beyond the single omnibus test. A formal mixed-effects analysis is planned for future work.

\section*{References}

\begin{enumerate}[label={[\arabic*]}, nosep, leftmargin=*]
    \item Bargh, J.~A., Chen, M., \& Burrows, L. (1996). Automaticity of
          social behavior: Direct effects of trait construct and stereotype
          activation on action. \textit{Journal of Personality and Social
          Psychology}, 71(2), 230--244.
    \item Batane, T. (2010). Turning to Turnitin to fight plagiarism among
          university students. \textit{Educational Technology \& Society},
          13(2), 1--12.
    \item Bilic-Zulle, L., Frkovic, V., Turk, T., Azman, J., \& Petrovecki,
          M. (2008). Is there an effective approach to deterring students
          from plagiarizing? \textit{Science and Engineering Ethics},
          14, 139--147.
    \item Brehm, J.~W. (1966). \textit{A Theory of Psychological Reactance}.
          Academic Press.
    \item Brooks, J.~L. (2012). Counterbalancing for serial order carryover
          effects in experimental condition orders. \textit{Psychological
          Methods}, 17(4), 600--614.
    \item Brown, T.~B., Mann, B., Ryder, N., et al. (2020). Language
          models are few-shot learners. \textit{Advances in Neural
          Information Processing Systems} 33. arXiv:2005.14165.
          \url{https://arxiv.org/abs/2005.14165}
    \item Corneille, O., \& Lush, P. (2023). Sixty years after Orne's
          American Psychologist article: A conceptual framework for
          subjective experiences elicited by demand characteristics.
          \textit{Personality and Social Psychology Review}, 27(1),
          83--101.
          \url{https://doi.org/10.1177/10888683221104368}
    \item Desaire, H., Chua, A.~E., Isom, M., Jarosova, R., \& Hua, D. (2023).
          Distinguishing academic science writing from humans or ChatGPT
          with over 99\% accuracy using off-the-shelf machine learning
          tools. \textit{Cell Reports Physical Science}, 4(6), 101426.
          \url{https://doi.org/10.1016/j.xcrp.2023.101426}
    \item Jiang, F., \& Hyland, K. (2025). Does ChatGPT argue like
          students? Bundles in argumentative essays.
          \textit{Applied Linguistics}, 46, 375--391.
          \url{https://doi.org/10.1093/applin/amae052}
    \item Karat, C.~M., Halverson, C., Horn, D., \& Karat, J. (1999).
          Patterns of entry and correction in large vocabulary continuous
          speech recognition systems. \textit{Proceedings of CHI '99},
          568--575.
    \item Krosnick, J.~A., Holbrook, A.~L., Berent, M.~K., Carson, R.~T.,
          Hanemann, W.~M., Kopp, R.~J., \ldots{} Conaway, M. (2002). The
          impact of ``no opinion'' response options on data quality.
          \textit{Public Opinion Quarterly}, 66(3), 371--403.
    \item Lee, K., Ippolito, D., Nystrom, A., Zhang, C., Eck, D.,
          Callison-Burch, C., \& Carlini, N. (2022). Deduplicating
          training data makes language models better. \textit{Proceedings
          of ACL 2022}. arXiv:2107.06499.
          \url{https://arxiv.org/abs/2107.06499}
    \item McCambridge, J., Witton, J., \& Elbourne, D.~R. (2014). Systematic
          review of the Hawthorne effect: New concepts are needed to study
          research participation effects. \textit{Journal of Clinical
          Epidemiology}, 67(3), 267--277.
    \item Mummolo, J., \& Peterson, E. (2019). Demand effects in survey
          experiments: An empirical assessment. \textit{American Political
          Science Review}, 113(2), 517--529.
    \item Noy, S., \& Zhang, W. (2023). Experimental evidence on the
          productivity effects of generative artificial intelligence.
          \textit{Science}, 381(6654), 187--192.
    \item Orne, M.~T. (1962). On the social psychology of the psychological
          experiment. \textit{American Psychologist}, 17(11), 776--783.
    \item Reviriego, P., Conde, J., Merino-Gomez, E., Martinez, G., \&
          Hernandez, J.~A. (2024). Playing with words: Comparing the
          vocabulary and lexical diversity of ChatGPT and humans.
          \textit{Machine Learning with Applications}, 18, 100602.
          \url{https://doi.org/10.1016/j.mlwa.2024.100602}
    \item Sandler, M., Choung, H., Ross, A., \& David, P. (2024). A
          linguistic comparison between human and ChatGPT-generated
          conversations. arXiv:2401.16587.
          \url{https://arxiv.org/abs/2401.16587}
    \item Schleimer, S., Wilkerson, D.~S., \& Aiken, A. (2003). Winnowing:
          Local algorithms for document fingerprinting. \textit{Proceedings
          of the 2003 ACM SIGMOD International Conference on Management
          of Data}, 76--85.
          \url{https://doi.org/10.1145/872757.872770}
    \item Schwarz, N. (1999). Self-reports: How the questions shape the
          answers. \textit{American Psychologist}, 54(2), 93--105.
          \url{https://doi.org/10.1037/0003-066X.54.2.93}
    \item Weingarten, E., et al. (2016). From primed concepts to action: A
          meta-analysis of the behavioral effects of incidentally presented
          words. \textit{Psychological Bulletin}, 142(5), 472--497.
    \item Zenker, F., \& Kyle, K. (2021). Investigating minimum text length
          for lexical diversity indices. \textit{Assessing Writing}, 47,
          100505.
\end{enumerate}

\newpage
\appendix
\section*{Appendix}

\subsection*{A. Full Assignment Table (PIDs 1 to 12)}

\begin{table}[H]
\centering
\small
\begin{tabular}{@{} c l l l @{}}
\toprule
\textbf{PID} & \textbf{Condition} & \textbf{Task 1} & \textbf{Task 2} \\
\midrule
 1 & Control & For     & Against \\
 2 & Test    & For     & Against \\
 3 & Control & Against & For \\
 4 & Test    & Against & For \\
 5 & Control & For     & Against \\
 6 & Test    & For     & Against \\
 7 & Control & Against & For \\
 8 & Test    & Against & For \\
 9 & Control & For     & Against \\
10 & Test    & For     & Against \\
11 & Control & Against & For \\
12 & Test    & Against & For \\
\bottomrule
\end{tabular}
\caption{Assignment table for the first 12 participant IDs, showing the
repeating four-participant cycle.}
\end{table}

\subsection*{B. Sensitivity Analysis}

The headline result (45.87\% control-as-human) was tested under
several filtering conditions to check robustness:

\begin{table}[H]
\centering
\small
\begin{tabular}{@{} l c c @{}}
\toprule
\textbf{Slice} & \textbf{n} & \textbf{Ctrl-as-human} \\
\midrule
All responses               & 1,999 & 45.87\% \\
Drop affected docs          & 1,831 & 45.71\% \\
Time filter only ($>$12s)   & 1,415 & 43.46\% \\
Drop affected + time filter & 1,298 & 43.45\% \\
\bottomrule
\end{tabular}
\caption{Sensitivity slices. ``Affected docs'' are the two documents
with known data-collection anomalies. The time filter removes
responses faster than 12 seconds (10s forced reveal + 2s minimum).
The effect strengthens under stricter filtering.}
\end{table}

\subsection*{C. Full Per-Document Chosen-as-Human Rates}

\begin{table}[H]
\centering
\scriptsize
\renewcommand{\arraystretch}{1.1}
\begin{tabular}{@{} l l r r r @{}}
\toprule
\textbf{Doc ID} & \textbf{Condition} & \textbf{Seen} & \textbf{Chosen} & \textbf{Rate} \\
\midrule
P26\_AGAINST & treatment & 102 & 78 & 76.5\% \\
P32\_AGAINST & treatment & 91 & 67 & 73.6\% \\
P26\_FOR     & treatment & 76 & 55 & 72.4\% \\
P40\_AGAINST & treatment & 92 & 64 & 69.6\% \\
P48\_FOR     & treatment & 82 & 56 & 68.3\% \\
P40\_FOR     & treatment & 77 & 50 & 64.9\% \\
P27\_AGAINST & control   & 109 & 68 & 62.4\% \\
P39\_FOR     & control   & 111 & 68 & 61.3\% \\
P27\_FOR     & control   & 108 & 65 & 60.2\% \\
P34\_FOR     & treatment & 92 & 55 & 59.8\% \\
\midrule
\multicolumn{5}{c}{\textit{(middle 21 documents omitted for space)}} \\
\midrule
P19\_FOR     & control   & 110 & 35 & 31.8\% \\
P44\_AGAINST & treatment & 82 & 25 & 30.5\% \\
P25\_AGAINST & control   & 114 & 34 & 29.8\% \\
\bottomrule
\end{tabular}
\caption{Top 10 and bottom 3 documents by chosen-as-human rate.
Treatment documents dominate the top of the table; control documents
dominate the bottom.}
\end{table}

\subsection*{D. Code and Data}

All analysis code, experiment data, and platform source code are available upon request. Please contact the author at dtabach3@gatech.edu.

\end{document}